\documentstyle[pre,aps,epsf,floats]{revtex}

\def\ggl{\hspace{0.5ex}\raisebox{0.4ex}{$\scriptstyle\geq$}\hspace{0.5ex}}
\def\kgl{\hspace{0.5ex}\raisebox{0.4ex}{$\scriptstyle\leq$}\hspace{0.5ex}}

\begin{document} 

\flushbottom

\draft
\twocolumn[\hsize\textwidth\columnwidth\hsize\csname @twocolumnfalse\endcsname

\title{Logarithmic corrections of the avalanche distributions
of sandpile models at the upper critical dimension}

\author{S. L\"ubeck\cite{SvenEmail} }

\address{
Theoretische Tieftemperaturphysik, 
Gerhard-Mercator-Universit\"at Duisburg,\\ 
Lotharstr. 1, 47048 Duisburg, Germany \\}

\date{Received 20 May 1998}

\maketitle

\begin{abstract}
We study numerically the dynamical properties 
of the BTW model on a square lattice for various dimensions.
The aim of this investigation is to determine
the value of the upper critical dimension
where the avalanche distributions are
characterized by the mean-field exponents.
Our results are consistent
with the assumption that the scaling behavior
of the four-dimensional BTW model
is characterized by the mean-field exponents
with additional logarithmic corrections.
We benefit in our analysis from the 
exact solution of the directed BTW model
at the upper critical dimension
which allows to derive how 
logarithmic corrections affect the
scaling behavior at the upper critical dimension.
Similar logarithmic corrections forms
fit the numerical data for the four-dimensional BTW model, 
strongly suggesting that the value of the upper critical
dimension is four.
\end{abstract}


]  

\setcounter{page}{1}
\markright{\rm
Phys. Rev.~E {\bf ?}, ? (1998), accepted for publication. }
\thispagestyle{myheadings}
\pagestyle{myheadings}

\section{Introduction}

The concept of self-organized criticality
introduced by Bak, Tang, and Wiesenfeld 
allows to describe scale invariance 
in driven systems~\cite{BAK}.
Sandpile models and especially the 
Bak-Tang-Wiesenfeld (BTW) sandpile model
are known as the paradigm of 
self-organized criticality.
The steady state dynamics of the system is characterized by
the probability distributions
for the occurrence of relaxation clusters of a certain  size, area,  
duration, etc.
Despite numerous theoretical
efforts~\cite{DHAR_2,MAJUM_1,PRIEZ_1,IVASH_1} the values of the exponents of the
probability distribution 
characterizing the critical behavior of the system 
were determined only numerically for $D=2$ and $D=3$~\cite{LUEB_2,LUEB_4}.
These investigations are based on an accurate finite-size
scaling analysis and were confirmed in a recently
published work~\cite{CHESSA_1}.
In higher dimensions the scaling behavior of the 
BTW model is still controversial.
Especially, the value of the upper critical dimension~$D_u$
where the mean-field solution describes the
scaling behavior of the system is not known exactly.
Whereas renormalization group approaches
predicted $D_u=4$~\cite{OBUKHOV_1,DIAZ_1,DIAZ_2} the results 
of numerical simulations are not consistent.
Several authors were led by their investigations 
to the conjecture that $D_u=4$~\cite{LUEB_4,GRASS_1}.
On the other hand comparable simulations 
in various dimension display no mean-field 
behavior for $D=4$ which was interpreted as
an evidence that the values of the upper critical
dimension is greater than four~\cite{CHESSA_1,CHRIS_2}.

In this paper we consider the BTW model in various 
dimensions and improve the accuracy of the analysis
significantly.
Our analysis reveals that the scaling behavior of
the four dimensional model is characterized 
by the mean-field exponents with additional
logarithmic corrections.
We benefit in our analysis from the 
exact solution of the directed BTW model
which displays logarithmic corrections
at the upper critical dimension~$D_u=3$~\cite{DHAR_1}.
This solution is used in order to develop a scaling 
analysis for the directed BTW model which takes these 
logarithmic corrections into account.
This type of scaling analysis will then be applied
to the usual BTW model.
The important result of this analysis is that
the scaling behavior of the probability distributions
as well as the usual finite-size scaling ansatz
are affected by logarithmic corrections for $D=4$.
These logarithmic corrections are a particular
feature of the four-dimensional system
and could not observed in higher dimensions.
This suggest that the value of the upper critical
dimension is indeed four.

\section{The BTW model}

We consider the $D$-dimensional BTW model on a square lattice
of linear size $L$ in which integer variables $E_{\bf r}\ggl 0$ represent
local energies.
One perturbes the system by adding particles at a randomly chosen site
${\bf r}$ according to
\begin{equation}
E_{\bf r} \, \mapsto \, E_{\bf r}+1\, .
\label{eq:perturbation}
\end{equation}
A site is called unstable if the corresponding  energy $E_{\bf r}$ 
exceeds a critical value $E_c$, i.e., if $E_{\bf r} \ggl E_c$, 
where $E_c$ is given by $E_c=2D$.
An unstable site relaxes, its energy is decreased by $E_c$
and the energy of the $2D$ next neighboring sites are increased by one unit, i.e.,
\begin{equation}
E_{\bf r}\;\to\;E_{\bf r}\,-\,E_c
\label{eq:relaxation_1}
\end{equation}
\begin{equation}
E_{{\rm nn},\bf r}\;\to\;E_{{\rm nn},\bf r}\;+\;1.
\label{eq:relaxation_2}
\end{equation}
In this way the neighboring sites may be activated and 
an avalanche of relaxation events may take place.
The sites are updated in parallel until all
sites are stable.
Starting with a lattice of randomly distributed energies 
$E \in\{0,1,2,...,E_c-1\}$ the system is perturbed according to
Eq.~(\ref{eq:perturbation})
and Dhar's 'burning algorithm' is applied in order to check if the 
system has reached the critical steady state \cite{DHAR_2}.
Usually one studies several different quantities in
order to characterize the avalanches:  
the number of relaxation events $s$ (size), 
the number of distinct toppled lattice site $a$ (area or volume), 
the duration $t$, and the
radius $r$.
In the critical steady state the corresponding probability 
distributions should obey power-law behavior 
characterized by exponents $\tau_s$, $\tau_a$,  $\tau_t$ and $\tau_r$
according to
\begin{equation}
P_x(x) \, \sim \, x^{-{\tau_x}},
\label{eq:prob_dist}
\end{equation}
with $x \in \{s,a,t,r\}$.
Because a particular lattice site may topple several times the 
number of toppling events exceeds the number of
distinct toppled lattice sites, i.e., $s \ggl a$.
It is known that multiple toppling events can
be neglected for $D \ggl 3$~\cite{LUEB_4,GRASS_1}, i.e.,~the
distributions $P_s(s)$ and $P_a(a)$
display the same scaling behavior and especially $\tau_s=\tau_a$.

Scaling relations for the exponents $\tau_s, \tau_a, \tau_t$ and
$\tau_r$ can be obtained if one assumes that the 
size, area, duration and radius scale as a 
power of each other, for instance 
\begin{equation}
t \, \sim \, r^{\gamma_{tr}}.
\label{eq:gam_tr}
\end{equation}
The transformation law of probability distributions 
$P_t(t) \mbox{d}t=P_r(r) \mbox{d}r$ leads to
the scaling relation
\begin{equation}
{\gamma_{tr}}\;=\;\frac{\tau_r-1}{\tau_t-1}.
\label{eq:gam_tau_tr}
\end{equation}
The scaling exponents~$\gamma_{xx'}$ are important for the
description of the avalanches properties and their
propagation. 
For instance the exponent $\gamma_{sa}$ indicates
if multiple toppling events are relevant ($\gamma_{sa}>1$)
or irrelevant ($\gamma_{sa}=1$).
Since the exponent $\gamma_{ar}$ determine
the scaling behavior of the avalanche area with
its radius, $\gamma_{ar}$~is an appropriate tool to investigate whether
the avalanche shape displays a fractal
behavior or not.
Finally, the exponent $\gamma_{tr}$ is usually identified
with the dynamical exponent~$z$.

The measurement of the 
probability distributions and the corresponding 
exponents [Eq.~(\ref{eq:prob_dist})]
is affected by the finite system size~$L$.
If the avalanche exponents
$\tau_x$ exhibit no system size dependence
the finite-size scaling analysis could 
be applied~\cite{KADANOFF}. 
In that case
the probability distributions 
obey the scaling equation
\begin{equation}
P_x(x,L) \; = \;
L^{-\beta_x} \, g_x ( x L^{-\nu_x} ), 
\label{eq:fss_simple}
\end{equation}
where the exponents have to fulfill
the scaling equation $\beta_x = \tau_x \nu_x$~\cite{KADANOFF}.
The exponent $\nu_x$ determines the cut-off 
behavior of the probability distribution
and it was shown that $\nu_x = \gamma_{xr}$
(see for instance~\cite{LUEB_4}).
The advantage of the finite size scaling analysis is that it 
yields additionally to the avalanche
exponents $\tau_x$ the important scaling exponents:~the 
avalanche dimension~$\nu_a$,
the dynamical exponent~$\nu_{t}=z$, etc.

The value of the upper critical dimension $D_u$ of 
the undirected BTW model is not known rigorously.
Several attempts were made to determine
the value of $D_u$ using numerical 
simulations~\cite{LUEB_4,CHESSA_1,GRASS_1,CHRIS_2}.
Usually one considers the probability 
distributions and compares the avalanche exponents
with the known mean-field 
values~(see for instance~\cite{VERGELES_1}).
But due to the limited computer power
the implementation of the higher dimensional
systems reduces considerably the system sizes~$L$
and consequently also the straight portion of the
probability distributions.
This makes a determination of the avalanche
exponents via regression very difficult for $D> 3$.
This disadvantage can be avoided by applying a 
finite-size scaling analysis. 
Our results obtained in this way are consistent 
with the assumption that~$D_u=4$ and that the
avalanche dimension is~$v_{a}=4$ for $D\ggl 4$~\cite{LUEB_4}.

Recently, Chessa~{\it et~al.}~considered the
BTW model in various dimensions using the same
finite-size scaling analysis~\cite{CHESSA_1}.
Compared to~\cite{LUEB_4} they examined larger
system sizes in $D\ggl 3$ and used an improved 
statistics~(up to $10^7$ nonzero avalanches).
From their results which differ for $D\ggl 4$
from those in~\cite{LUEB_4} they concluded that $D=4$
is not the upper critical dimension.
Especially they obtained from their
finite-size scaling analysis $v_{a}\approx 3.5$
for the four-dimensional BTW model, i.e.,~the
avalanches display a fractal behavior already
for $D=4$.
The origin of these conflicting results is that 
the used statistics ($2\times 10^6$ nonzero avalanches) 
in~\cite{LUEB_4} is not sufficient.
Especially the fluctuating data points at the cut-off
of the distribution~$P_a(a)$ lead to 
uncertain results~(see Fig.~5 in~\cite{LUEB_4}).
For instance it is possible to obtain
with this data a collapse of the distributions $P_a(a,L)$ for values
of the avalanche dimension between $v_a=3.4$ and $v_a=4.1$.

Thus, there is no agreement in the literature
on the behavior 
of the BTW model in different 
dimensions:
Chessa~{\it et~al.}~concluded from their analysis
that the upper critical dimension is larger than four
and that the avalanches displays fractility
already for $D=4$.
On the other hand there exist several theoretical
approaches which leads to the conclusion that
$D_u=4$:~Real 
space~\cite{OBUKHOV_1} as well as momentum 
space~\cite{DIAZ_1,DIAZ_2} renormalization group 
analysis predicted both $D_u=4$.
From their exact solution of the BTW model
on the Bethe lattice Majumdar and Dhar concluded 
that $D_u \ggl 4$ because 
the fractal dimension of avalanche clusters must be
lower than that of the embedding space~\cite{DHAR_3}.
This leads the authors to the conjecture that the avalanches
are compact for~$D\kgl D_u$
and fractal above the critical dimension.
This fractility of the avalanche
structure was already observed.
Considering the avalanche propagation in
higher dimension it was found that the avalanches 
are characterized
by a compact activation front for $D=3$ and $D=4$.
For $D>4$ the compact shape of the activation 
front is lost and several branches
propagate through the system without
coalesce together again~(see Fig.~8 in~\cite{LUEB_4}).
Here, the avalanche propagation can be described 
as a branching process which is the main feature
of the mean-field solution of sandpile 
models~(see for instance~\cite{VERGELES_1}).
Assuming that the clusters are compact and neglecting
multiple toppling events (which is justified 
for $D=3$ and $D=4$~\cite{LUEB_4,GRASS_1}) Zhang derived in the
continuum limit the equation $\tau_a=2-2/D$~\cite{ZHANG_1}
which gives the mean-field value $\tau_a=3/2$ again for~$D=4$.

This incoherent picture of the behavior 
of the BTW model in higher dimensions leads
us to reconsider the avalanche distributions again
and compare our results with those of~\cite{CHESSA_1}. 
In contrast to our previous work~\cite{LUEB_4} we use now larger 
system sizes ($L\kgl 128$ for $D=4$,
$L\kgl 48$ for $D=5$, and 
$L\kgl 24$ for $D=6$) and increase
the statistics significantly, i.e.,~we averaged
all measurements over at least $5\times 10^7$
non-zero avalanches.
As usual we measure the avalanche distributions
[Eq.~(\ref{eq:prob_dist})] by counting the numbers
of avalanches corresponding to a given area, duration, etc.
and integrate these numbers over bins of increasing
length (see for instance~\cite{MANNA_2}).
Successive bin length increases by a factor $b>1$. 
Throughout this work we performed all measurements
with the factor $b=1.2$ since larger values
of $b$ may change the cut-off shape of the 
distributions.
Applying the finite-size scaling analysis
this effect could lead to uncertain 
results for the scaling exponent $v_a$ (we found
that this effect has to be taken into consideration
at least for $b\ggl 1.5$).

\begin{figure}[b]
 \epsfysize=6.8cm
 \epsffile{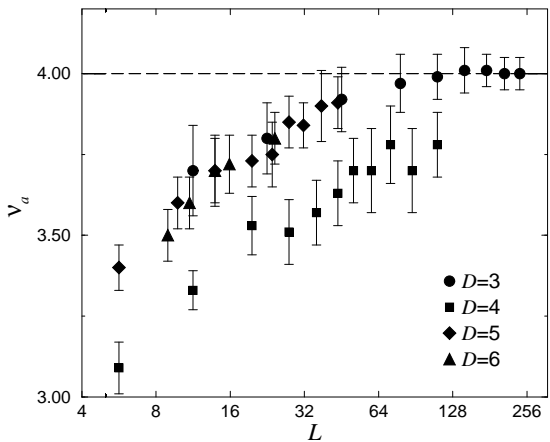}
 \caption{The finite-size scaling exponents~$\nu_a$ of the
	  BTW model for various dimensions.
          The values of the exponents are obtained from the
	  finite-size scaling analysis [Eq.~(\ref{eq:fss_simple})] 
	  of two probability distributions corresponding
	  to two different system size~$L_1$ and $L_2$
	  and are plotted as a function of $L=(L_1 L_2)^{1/2}$.
	  In order to compare the different dimensions $\nu_a+1$
          is plotted for $D=3$.
 \label{fig:BTW_avalanche_dimension}} 
\end{figure}

We focus our attention on the finite-size
scaling analysis.
Performing this analysis
it is informative to produce the data-collapse not only
for all curves corresponding to different
system sizes but also to check the obtained
data collapse for selected curves.
For instance the finite-size scaling analysis
of two curves corresponding to two 
successive system size ($L_1<L_2$) allows to
check whether the actual scaling regime
is already reached.
This analysis is shown in Fig.~\ref{fig:BTW_avalanche_dimension}
for the three-dimensional BTW model
where it is known that finite-size scaling 
works for $L\ggl 64$~\cite{LUEB_4}.
If one performs the finite-size scaling 
analysis for two system sizes with
$L_1<L_2<64$ it is possible to obtain
a data collapse (with small but systematic deviations, especially
at the cut-off) but then the scaling
exponents depend on the system sizes.
In Fig.~\ref{fig:BTW_avalanche_dimension} we plot
the scaling exponent $\nu_a(L_1,L_2)$ as a function
of the average system size $L=(L_1 L_2)^{1/2}$.
With increasing system sizes the exponent tends to the
value $\nu_a=3$.
For $L \ggl 64$ no significant system size dependence
could be observed, i.e., a cross-over to the actual
scaling regime where finite-size scaling works
takes place at $L_{co} \approx 64$.

Analogous to the three-dimensional model we
perform the same analysis for $D=4$, $D=5$, and $D=6$
and plot the obtained results 
in Fig.~\ref{fig:BTW_avalanche_dimension}.
It is remarkable that within the error-bars 
the values $\nu_a(D=3)+1$, $\nu_a(D=5)$, and 
$\nu_a(D=6)$ display for small system sizes 
the same finite-size dependence whereas the 
behavior of the four-dimensional system differs
significantly from the other dimensions.
The conjecture that the system size dependence of $\nu_a$ is
independent of the dimension (except of the case $D=4$) 
implies that the cross-over to the actual scaling regime
takes place at a comparable value $L_{co} \approx 64$.
This could explain why the finite-size
scaling analysis performed by Chessa~{\it et al.}~for
$D\ggl 5$ yields exponents which are lower than the mean-field 
value $\nu_a=4$~\cite{CHESSA_1}.
Their considered system sizes for $D\ggl 5$ are
outside the scaling regime where finite-size scaling
works.
Thus, they (and of course all other previous numerical 
investigations~\cite{LUEB_4,GRASS_1,CHRIS_2}) 
observed only the cross-over to
the real scaling regime and not the real scaling 
behavior itself.

The significantly different behavior of the scaling 
exponent $\nu_a$ for $D=4$~(see Fig.~\ref{fig:BTW_avalanche_dimension})
is remarkable since with increasing system size 
no cross-over to a scaling regime
with a system size independent exponent~$\nu_a$
could be observed.
It seems that the scaling behavior of the 
four-dimensional model differs in principle
from all other dimensions.
A possible explanation is that the value of
the upper critical dimension is~$D_u=4$.
Then the unique behavior of the exponent~$\nu_a$ and
the observed deviations to the expected pure mean-field 
scaling behavior for $D=4$~\cite{CHESSA_1} 
could be explained by additional
logarithmic corrections which affect
the scaling behavior and which typically
occur at the upper critical dimension.

In the rest of this paper we
will show that our results are consistent
with the assumption that the scaling behavior
of the four-dimensional BTW model
is characterized by the mean-field exponents
with additional logarithmic corrections.
In the next section we consider 
the BTW model with a preferred direction
of the dynamics.
This directed BTW model is exactly solved
and it is known that logarithmic corrections occur
for $D_c=3$~\cite{DHAR_1}.
The directed BTW model is therefore a suitable 
paradigm to learn how
the logarithmic corrections enters the scaling behavior
at the upper critical dimension.
This method of analysing 
will then be applied to the four dimensional 
BTW model in the last section.

\section{The directed BTW model at the upper
critical dimension}

In this section we consider the directed version of 
the BTW model which was introduced and exactly solved
in all dimensions by Dhar and Ramaswamy~\cite{DHAR_1}.
Directed models are characterized
by a preferred direction of the toppling rules.
For instance a relaxation process takes place in a
two-dimensional model if the energy of a given lattice site $(i,j)$
exceeds the critical value $E_c=D$: 
\begin{eqnarray}
E_{i,j} \;   & \to & \; E_{i,j} \,- \, E_c \nonumber \\ \nonumber\\
E_{i+1,j} \; & \to & \; E_{i+1,j} \, + \, E_c/D\\ \nonumber\\
E_{i,j+1} \; & \to & \; E_{i,j+1} \, + \, E_c/D\nonumber.
\label{eq:anisotope_relax}
\end{eqnarray}
One usually considers in simulations directed systems 
on a square lattice with periodic boundary conditions in the
direction perpendicular to the preferred direction
and open boundary conditions parallel to the 
preferred direction.
The system is perturbed on the first line 
only (top of the pile) and particle could leave
the system only on the last line (bottom of the pile).

Caused by the definition of the toppling rules no 
multiple toppling events can 
occur~($\Rightarrow \tau_s=\tau_a$).
Since the perturbation takes place only on the 
top of the pile the average flux of particles
through a surface in a given distance from the top
is constant.
This flux conservation leads to the scaling
relation~\cite{DHAR_1} 
\begin{equation}
\tau_a \; = \; 2 - \frac{1}{\tau_t}. 
\label{eq:tau_a_tau_t}
\end{equation}
According to Dhar and Ramaswamy the avalanche exponents 
of the two-dimensional model can be obtained by mapping 
the avalanche propagation onto a random walk  and
one gets $\tau_a=4/3$ and $\tau_t=3/2$, respectively.

\begin{figure}[t]
 \epsfxsize=8.6cm
 \epsfysize=8.6cm
 \epsffile{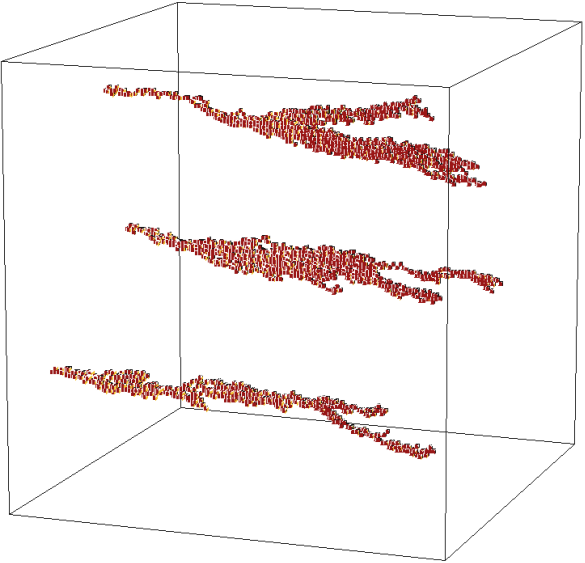}
 \caption{Snapshots of three arbritrary chosen avalanche 
	  of the directed BTW model at the upper
	  critical dimension for $L=128$.
 \label{fig:Dhar_avalanches_3D}} 
\end{figure}

For $D\ggl 3$ the exponents equal the mean-field
values, i.e.,~$\tau_a=3/2$ and $\tau_t=2$, and additional logarithmic 
corrections to the power-law behavior occur in 
$D=3$ which is the value of the
upper critical dimension~\cite{DHAR_1}.
A snapshot of several avalanches
of the three-dimensional model are
shown in Fig.~\ref{fig:Dhar_avalanches_3D}.
The shape of the avalanches reminds of  a
branching process which characterizes the 
avalanche propagation in the mean-field
solution.

\begin{figure}[t]
 \epsfysize=6.8cm
 \epsffile{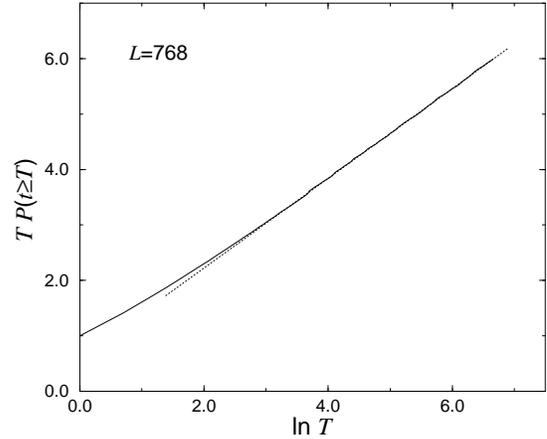}
 \caption{The probability distributions $P(t\protect\ggl T)$ of an 
	  avalanche of duration greater than
          or equal to $T$ for the directed
	  BTW model at the upper critical dimension~$D_u=3$.
	  According to Eq.~(\ref{eq:prob_dis_integrated})
	  $T\, P(t\protect\ggl T) $ is
	  plotted as a function of $\ln{T}$.
	  The dotted line is plotted
	  to guide the eye.
 \label{fig:dhar_3d_duration_int}} 
\end{figure}

\begin{figure}[t]
 \epsfysize=6.8cm
 \epsffile{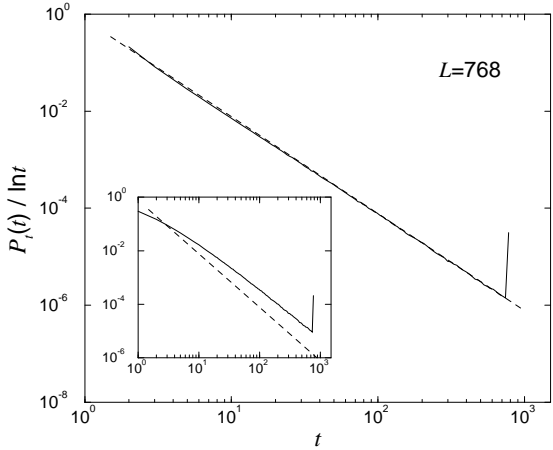}
 \caption{The probability distributions $P_t(t)$ of the directed
	  BTW model for $D=3$.
	  According to Eq.~(\ref{eq:prob_dist_duration_dhar_3d})
	  $P_t(t) / \ln{t}$ is
	  plotted as a function of the duration $t$.
	  The dashed line corresponds to a power-law
	  with the exponent $\tau_t=2$.
	  In the inset we plot $P_t(t)$ vs. $t$. 
	  Here, the curvature is caused by the logarithmic 
	  corrections. 
	  The inset shows that wrong results of the exponents
	  were obtained if one does not take the logarithmic
	  corrections into account.
 \label{fig:dhar_3d_duration}} 
\end{figure}

According to the exact solution of Dhar and Ramaswamy 
the mean square flux $m(t)$ out of a given surface $t$
is given by 
\begin{equation}
m(t) \; = \; \sum_{t'=1}^{t} \, F(t')
\label{eq:dhar_mean_flux_01}
\end{equation}
with $F(t) \sim 1/\ln{t}$ for $D=3$~\cite{DHAR_1}.
Since the average flux through a surface $t$
is constant in the steady state the probability
distribution of an avalanche of duration greater than
or equal to $T$ scales in leading order as
\begin{equation}
P(t \ggl T) \; \sim \; \frac{1}{m(T)} \; \sim \; \frac{\ln{T}}{T}.
\label{eq:prob_dis_integrated}
\end{equation}
The corresponding plot of the rescaled
distribution $T\, P(t \ggl T)$
as function of the duration in a logarithmic
diagram is shown in Fig.~\ref{fig:dhar_3d_duration_int}
and confirms Eq.~(\ref{eq:prob_dis_integrated}).
The scaling behavior of the corresponding
density distribution~$P_t(t)$ is then given by
\begin{equation}
P_t(t) \; \sim \; \frac{\ln{t}}{t^2}.
\label{eq:prob_dist_duration_dhar_3d}
\end{equation}
In Fig.~\ref{fig:dhar_3d_duration} we plot
the rescaled distribution $P_t(t) / \ln{t}$
as a function of the duration $t$.
The rescaled distribution exhibits a power-law
behavior with the exponent $\tau_t=2$, in 
agreement with Eq.~(\ref{eq:prob_dist_duration_dhar_3d}).
The inset of Fig.~\ref{fig:dhar_3d_duration}
shows that a fit of the unscaled 
distribution~$P_t(t)$ leads to lower
values of the exponents~$\tau_t$,
i.e.,~a simple regression analysis can lead to the wrong
result that the probability distributions
are not characterized by the mean-field exponents.

\begin{figure}[t]
 \epsfysize=6.8cm
 \epsffile{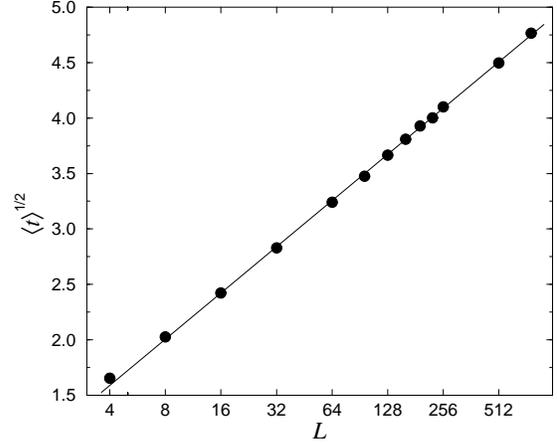}
 \caption{The square root of the average avalanche 
	  distribution ${\langle t \rangle_L^{1/2}}$	
	  vs.~$L$ of the 
	  directed BTW model for $D=3$.
	  The solid line corresponds to a logarithmic
          dependence of $\langle t \rangle_L^{1/2}$
	  according to Eq.~(\ref{eq:aver_duration}).
 \label{fig:dhar_3d_aver_duration}} 
\end{figure}

The scaling behavior of the average avalanche 
duration~$\langle t \rangle_L$ confirms the
relevance of the logarithmic corrections.  
Using Eq.~(\ref{eq:prob_dist_duration_dhar_3d}) the 
average duration is given by 
\begin{equation}
\langle t \rangle_L \; = \;
\int\limits^{t_{\rm max}}t \, P_t(t) \, \mbox{d}t
\; \sim \;
(\ln{L})^2 ,
\label{eq:aver_duration}
\end{equation}
because in directed models the maximum value
of the avalanche duration $t_{\rm max}$ equals the system size $L$.
The scaling behavior of the average duration 
clearly displays the relevance of the 
logarithmic corrections since
without these corrections the average duration
scales as $\sim \ln{L}$.
In order to confirm this result
we plot in Fig.~\ref{fig:dhar_3d_aver_duration} 
the square root of the average duration
as a function of the system size $L$ in a logarithmic
diagram.
The scaling behavior of the average
duration agrees with Eq.~(\ref{eq:aver_duration}).

The scaling behavior of the probability 
distribution~$P_a(a)$ of the avalanche area
displays also logarithmic corrections.
The area $a$ of an avalanche of total duration~$t$ is
determined by the average number of toppling
events in each surface $t'\kgl t$ (see~\cite{DHAR_1})
and one gets to leading order
\begin{equation}
a(t) \; = \; 
\sum_{t'=1}^t \, m(t) 
\; \sim \;
\frac{t^2}{\ln{t}} .
\label{eq:area_duration_dhar_3d}
\end{equation}
Instead of the usual scaling behavior $a \sim t^2$
which is valid for $D>D_u$ the leading order of the area scales with
the duration as \mbox{$a \sim t^2/\ln{t}$}.
Since the maximum value of the duration~$t_{\rm max}$ equals
the system size the maximum avalanche area scales as 
\begin{equation}
a_{\rm max} \; \sim \;
\frac{L^2}{\ln{L}}.
\label{eq:max_area_L}
\end{equation}
The maximum area $a_{\rm max}$ determines the cut-off
behavior of the probability distribution and 
Eq.~(\ref{eq:max_area_L}) indicates 
that the usual finite-size scaling 
ansatz~[Eq.~(\ref{eq:fss_simple})]
has to be modified in the presence of 
logarithmic corrections.
In the following we derive this modified
finite-size scaling ansatz which
describes the scaling behavior of the
avalanche distribution~$P_a(a)$ for $D=D_u$.
We assume that the leading order of the 
probability distribution
of the avalanche area is given by
\begin{equation}
P_a(a) \; \sim \;
\frac{(\ln{a})^{x_a}}{a^{\tau_a}}
\label{eq:prob_dist_area_dhar_3d}
\end{equation}
with the mean-field exponent $\tau_a=3/2$ 
and where the exponent of the 
logarithmic corrections $x_a$ has to be determined. 
Comparing Eq.~(\ref{eq:prob_dist_duration_dhar_3d}) with
Eq.~(\ref{eq:prob_dist_area_dhar_3d}) the 
corresponding exponent of the duration distribution
is given by $x_t=1$.
Using the transformation law for probability
distributions $P_a(a) {\rm d}a=P_t(t) {\rm d}t$ one can
derive the exponent $x_a$.
Inserting Eq.~(\ref{eq:area_duration_dhar_3d}) into 
Eq.~(\ref{eq:prob_dist_area_dhar_3d}) one gets
\begin{equation}
P_a[a(t)] \; \sim \;
t^{-3} \, (\ln{t})^{x_a+3/2} \,
\left ( 2 - \frac{\ln{\ln{t}}}{\ln{t}} \right)^{x_a}
\label{eq:prob_dist_area_with_duration}
\end{equation}
and analogous
\begin{equation}
\frac{{\rm d}a(t)}{{\rm d}t} \; \sim \;
t \, \frac{2 \ln{t} -1}{(\ln{t})^2}.
\label{eq:prob_trans_02}
\end{equation}
The term of leading order in the 
transformation law has to vanish and
thus we get $x_a=1/2$.

\begin{figure}[t]
 \epsfysize=6.8cm
 \epsffile{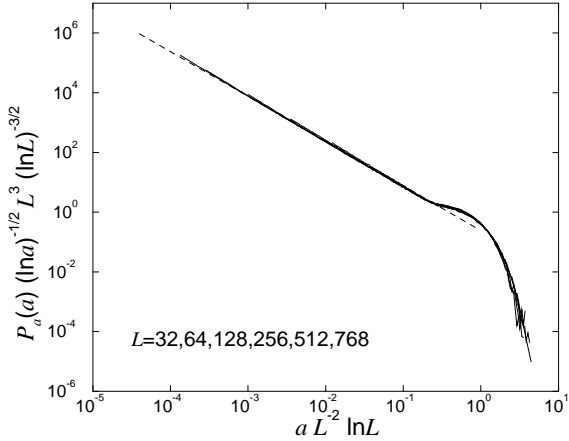}
 \caption{The modified finite-size scaling plot
	  of the probability distribution~$P_a(a)$
	  of the directed BTW model for~$D=3$.
          The data collapse of the different curves
	  corresponding to different system sizes~$L$
          confirms~Eq.~(\ref{eq:modified_fss_02}).
	  The dashed line corresponds to a power-law
	  with the mean-field exponent $\tau_a=3/2$.
 \label{fig:dhar_3d_area_fss}} 
\end{figure}

Due to the logarithmic corrections of the 
probability distribution $P_a(a)$ the
simple finite-size scaling ansatz 
Eq.~(\ref{eq:fss_simple})
does not work.
The simplest ansatz is to assume that
the rescaled distribution $P_a(a) (\ln{a})^{x_a}$
obeys the finite-size scaling equation
\begin{equation}
P_a(a,L) \, (\ln{a})^{-x_a} 
\; \sim \;
f(L) \; g ( a / a_{\rm max}  ) 
\label{eq:modified_fss_01}
\end{equation}
with the universal function $g$ and 
where the scaling function $f(L)$ has to be determined.
For low values of the argument of the universal function
($a \ll a_{\rm max}$) the rescaled probability distribution
is independent of the system size and characterized
by the power-law behavior $g(x)\sim x^{-\tau_a}$ only.
Thus we obtain $f(L) \sim a_{\rm max}^{-\tau_a}$.
Using the known scaling behavior of $a_{\rm max}$
we get the modified finite-scaling ansatz
\begin{equation}
P_a(a,L) \, (\ln{a})^{-x_a} 
\; = \;
L^{-2 \tau_a} \, (\ln{L})^{\tau_a}\,
\; g ( a L^{-2} \ln{L}    ) .
\label{eq:modified_fss_02}
\end{equation}
We present the corresponding scaling plot
in Fig.~\ref{fig:dhar_3d_area_fss}.
The data collapse of the different curves
corresponding to different system size~$L$
confirms the above analysis.

In summary we showed
that the scaling behavior of the 
directed BTW model at the upper critical dimension
is characterized by strong logarithmic
corrections.
These logarithmic corrections affect
the usual probability distributions~[Eq.~(\ref{eq:prob_dist})], 
the scaling equations [Eq.~(\ref{eq:gam_tr})],
and the finite-size scaling analysis~[Eq.~(\ref{eq:fss_simple})].
The corrections are relevant in the sense
that one has to take them into account
in order to describe the real scaling behavior,
otherwise one gets wrong values for the
exponents~(see Fig.~\ref{fig:dhar_3d_duration}).

Additionally we simulated
the directed BTW model for $D=4$
and performed a finite-size scaling analysis.
In agreement with the exact solution of Dhar and
Ramaswamy the simple finite-size scaling
ansatz, i.e.,~without logarithmic corrections,
works quite well and the corresponding
exponents equal the mean-field exponents.
Thus, the logarithmic corrections to the
scaling behavior occur only at the upper critical
dimension~$D_c=3$~\cite{DHAR_1}.

\section{The undirected BTW model for $D=4$}

In the following we return to the investigation
of the undirected BTW model
for $D=4$ and show that the avalanche distributions
are characterized by logarithmic corrections
comparable to the directed BTW model at the 
upper critical dimension.
First we generalize the scaling equations by
introducing certain exponents which describe 
the logarithmic corrections.
Guided by our previous analysis
we assume that the probability distributions
are given by
\begin{equation}
P_t(t) \; \sim \;
\frac{(\ln{t})^{x_t}}{t^{\tau_t}}
\quad \mbox{\rm and} \quad
P_a(a) \; \sim \;
\frac{(\ln{a})^{x_a}}{a^{\tau_a}}.
\label{eq:prob_dist_btw_4d}
\end{equation}
The maximal avalanche duration and area
which determine the cut-off behavior 
of the corresponding distributions
should scale with the system size as
\begin{equation}
t_{\rm max} \; \sim \;
\frac{L^{\nu_t}}{(\ln{L})^{N_t}}
\quad \mbox{\rm and} \quad
a_{\rm max} \; \sim \;
\frac{L^{\nu_a}}{(\ln{L})^{N_a}}.
\label{eq:maximal_btw_4d}
\end{equation}
The fifth introduced exponent describes
how the area scales with the duration
\begin{equation}
a(t) \; \sim \;
\frac{t^{\gamma_{at}}}{(\ln{t})^{\Gamma_{at}}}.
\label{eq:area_duration_btw_4d}
\end{equation} 
At the upper critical dimension 
the avalanche and scaling exponents 
equal the mean-field values $\tau_a=3/2$, $\tau_t=2$, 
$\nu_a=4$, $\nu_t=2$, and $\gamma_{at}=2$.
In this way the logarithmic corrections are
determined by five non-negative exponents
which have to fulfill two scaling relations.
The transformation law of probability distributions
leads to the first scaling equation:
\begin{eqnarray}
P_a[a(t)] \, \frac{{\rm d}a(t)}{{\rm d}t} \, {\rm d}t
\; & = & \; P_t(t) \, {\rm d}t \nonumber \\ 
 \Longrightarrow \quad \quad \quad   \frac{\Gamma_{at}}{2} \; & = & \; x_t \, - \, x_a, 
\label{eq:log_scal_rel_01} 
\end{eqnarray}
where we make use of the equation $\gamma_{at} = (\tau_t-1)/(\tau_a-1)$
and assume that the term of leading order of the logarithmic
corrections has to vanish.
Under the condition that the scaling behavior 
of the leading order of the maximum avalanche 
area $a_{\rm max}$ is given by Eq.~(\ref{eq:maximal_btw_4d}) 
and Eq.~(\ref{eq:area_duration_btw_4d})
we obtain the second scaling relation
\begin{eqnarray}
a_{\rm max} \; & = & \; a(t_{\rm max}) \; 
= \; \frac{L^{\nu_t \gamma_{at}}}{(\ln{L})^{N_t \gamma_{at} + \Gamma_{at}}}
 \nonumber \\ 
 \Longrightarrow \quad \quad \quad   N_a \; & = & \; \Gamma_{at} \, + \, \gamma_{at}
\, N_t.
\label{eq:log_scal_rel_02} 
\end{eqnarray}
Here, we use that the scaling exponents equal
the avalanche dimension ($\nu_a = \gamma_{ar}$) and 
the dynamical exponent ($\nu_t = \gamma_{tr}$),
respectively.
The relation $\gamma_{tr} \gamma_{at} = \gamma_{ar}$~\cite{CHRIS_2}
leads to Eq.~(\ref{eq:log_scal_rel_02}). 
Thus, the logarithmic corrections to the
usual scaling behavior
are determined by only three independent
exponents.

Corresponding to the directed BTW model for \mbox{$D=D_u$}
we assume that the distributions obey the
finite-size scaling ansatz
\begin{equation}
P_t(t,L) \, (\ln{t})^{-x_t} 
\; \sim \;
f_t(L) \; g_t ( t / t_{\rm max}  ), 
\label{eq:modified_fss_btw_duration}
\end{equation} 
\begin{equation}
P_a(a,L) \, (\ln{a})^{-x_a} 
\; \sim \;
f_a(L) \; g_a ( a / a_{\rm max}  ) ,
\label{eq:modified_fss_btw_area}
\end{equation} 
where the scaling functions $f_t$ and $f_a$
are given by
\begin{equation}
f_t(L) 
\; \sim \;
L^{-\nu_t \tau_t} \; (\ln{L})^{N_t \tau_t},
\label{eq:modified_fss_btw_duration_f}
\end{equation} 
\begin{equation}
f_a(L) 
\; \sim \;
L^{-\nu_a \tau_a} \; (\ln{L})^{N_a \tau_a},
\label{eq:modified_fss_btw_area_f}
\end{equation} 
since for small values of the argument
of the universal functions ($t\ll t_{\rm max}$
and $a\ll a_{\rm max}$) the probability
distributions are independent of the system size.
For $D=D_u$ the avalanche and scaling exponents 
equal the mean-field values
$\tau_a=3/2$, $\tau_t=2$, $\nu_a=4$, and $\nu_t=2$.
Thus the finite-size scaling analysis of the
area and duration distribution gives the 
correction exponents $N_t$, $x_t$, $N_a$, $x_a$,
and the obtained values have to fulfill 
the equation [Eq.~(\ref{eq:log_scal_rel_01}) and
Eq.~(\ref{eq:log_scal_rel_02})]
\begin{equation}
2(x_t-x_a) \; = \; N_a \, - \, 2 N_t .
\label{eq:log_scal_rel_03}
\end{equation}

\begin{figure}[t]
 \epsfysize=6.8cm
 \epsffile{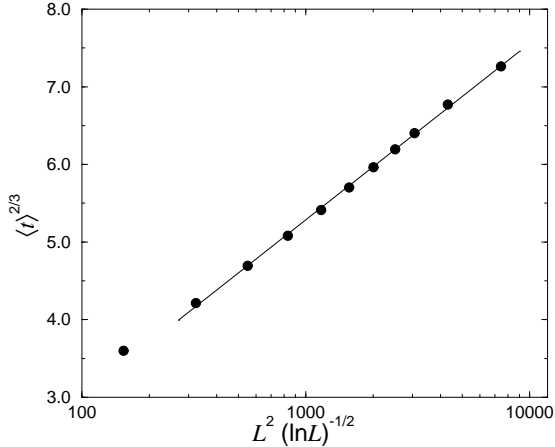}
 \caption{The system size dependence of the 
	  average avalanche 
	  duration ${\langle t \rangle_L}$ of the 
	  BTW model for $D=4$.	
	  To guide the eye we plot the solid line	
	  which corresponds to Eq.~(\ref{eq:aver_dura_btw_4d_log}).
 \label{fig:btw_4d_aver_duration}} 
\end{figure}
\begin{figure}[b]
 \epsfysize=6.8cm
 \epsffile{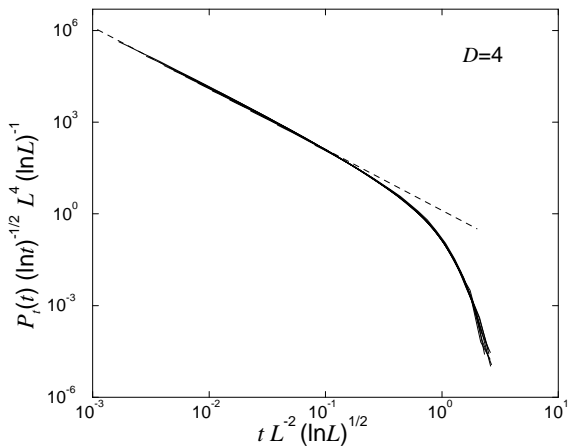}
 \caption{The modified finite-size scaling plot
	  of the probability distribution~$P_t(t)$
	  of the BTW model for~$D=4$ and 
	  $L=24,32,40,48,56,64,72,80,96,128$.
          The data collapse of the different curves
	  corresponding to different system sizes~$L$
          confirms~Eq.~(\ref{eq:modified_fss_btw_duration}).
	  The dashed line corresponds to a power-law
	  with the mean-field exponent $\tau_t=2$.
 \label{fig:btw_4d_dura_fss}} 
\end{figure}

Analogous to the analysis of the directed BTW model 
we consider the scaling behavior of the 
average avalanche duration $\langle t \rangle_L$ 
before we apply the finite-size scaling analysis.
According to Eq.~(\ref{eq:prob_dist_btw_4d}) and 
Eq.~(\ref{eq:maximal_btw_4d})
the average duration is given by
\begin{equation}
\langle t \rangle_L \; = \;
\int\limits^{t_{\rm max}}t \, P_t(t) \, \mbox{d}t
\; \sim \;
\left ( \ln{\frac{L^2}{(\ln{L})^{N_t}}} \right )^{x_t+1}
\label{eq:aver_dura_btw_4d_log}
\end{equation}
which allows additionally to the finite-size scaling analysis
an independent determination of the 
exponents $x_t$ and $N_t$.
We tried several values of $x_t$ and $N_t$ and obtained 
a good result for $x_t \approx 1/2$ and
$N_t \approx 1/2$.
In Fig.~\ref{fig:btw_4d_aver_duration} we present in an
logarithmic diagram $\langle t \rangle^{2/3}$
vs.~$L^2/(\ln{L})^{1/2}$. 
The plotted values are located on a straight line,
in agreement with Eq.~(\ref{eq:aver_dura_btw_4d_log}).

These values are confirmed by the finite-size
scaling analysis of the duration distribution
$P_t(t)$ according to Eq.~(\ref{eq:modified_fss_btw_duration}).
A satisfying data collapse is obtained for
$x_t=1/2$ and $N_t=1/2$ as one can see in
Fig.~\ref{fig:btw_4d_dura_fss}.

\begin{figure}[t]
 \epsfysize=6.8cm
 \epsffile{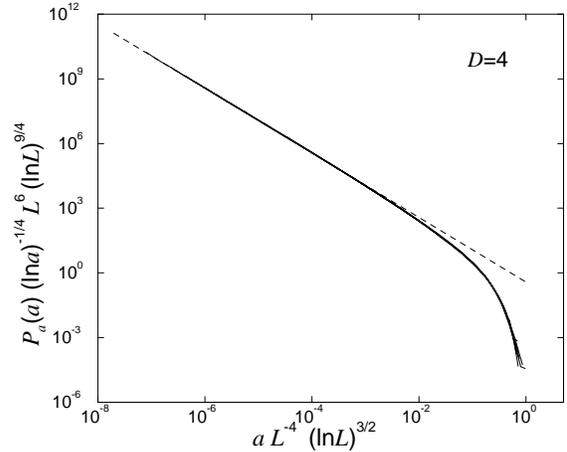}
 \caption{The modified finite-size scaling plot
	  of the probability distribution~$P_a(a)$
	  of the BTW model for~$D=4$ and
	  $L=24,32,40,48,56,64,72,80,96,128$.
          The data collapse of the different curves
	  corresponding to different system sizes~$L$
          confirms~Eq.~(\ref{eq:modified_fss_btw_area}).
	  The dashed line corresponds to a power-law
	  with the mean-field exponent $\tau_a=3/2$.
 \label{fig:btw_4d_area_fss}} 
\end{figure}

Now we consider the finite-size scaling analysis
of the area duration~$P_a(a)$.
Since the correction exponents have to fulfill
Eq.~(\ref{eq:log_scal_rel_03}) it must be possible 
to produce the data collapse of~$P_a(a)$ by varying
one parameter only 
if we assume that the above determined values $x_t=1/2$
and $N_t=1/2$ are correct.
We eliminated $N_a$ in the scaling ansatz 
[Eq.~(\ref{eq:modified_fss_btw_area}) and 
Eq.~(\ref{eq:modified_fss_btw_area_f})] 
and varied the exponent $x_a$.
An almost perfect data collapse 
is obtained for $x_a \approx 1/4$.
The corresponding scaling plot is shown
in Fig.~\ref{fig:btw_4d_area_fss} and 
confirms the accuracy of the above determined
exponents $x_t$ and $N_t$.
In contrast to the three dimensional model,
where the simple finite-size scaling works
for $L \ggl 64$~\cite{LUEB_4}, 
the finite-size behavior of the four-dimensional
model are governed by the logarithmic
corrections and the modified finite-size
scaling ansatz works very well already for $L \ggl 24$.
Finally we mention that Chessa~{\it et~al.}~who
used the simple finite-size scaling
ansatz obtained a less accurate data collapse 
for $L\ggl 48$~\cite{CHESSA_1} 
since they did not take the logarithmic corrections
into account.

\section{Conclusions}

We studied numerically the dynamical properties 
of the BTW model on a square lattice for $D\ggl 3$.
Our investigation of the avalanche distribution
which includes a careful examination of 
the finite-size corrections shows that 
analyses~\cite{LUEB_4,CHESSA_1} of the BTW model 
for $D\ggl 4$ are not conclusive.
Our results are consistent
with the assumption that the scaling behavior
of the four-dimensional BTW model
is characterized by the mean-field exponents
with additional logarithmic corrections.
We provide numerical tests for the theoretically
predicted logarithmic correction terms for the 
directed BTW model at the upper critical dimension 
$D_u=3$.
We introduce a refined finite-size scaling analysis
which takes these logarithmic 
corrections into account.
These logarithmic corrections occur
for $D=4$ only, strongly suggesting
that the value of the upper critical
dimension is four.
To proof this definitely in our opinion
it is necessary to
show that the distributions of the five
dimensional model are characterized by the pure
mean-field values.
Unfortunately, due to the limited computer power 
it is at present impossible to consider 
the actual scaling regime for $D=5$.

\acknowledgments
I would like to thank A.~Hucht and K.~D.~Usadel for
many helpful discussions and for a critical reading of 
the manuscript.
Finally, I wish to thank D.~Dhar
for useful discussions and his comments on the manuscript.
This work was supported by the
Deutsche Forschungsgemeinschaft through
Sonderforschungsbereich 166, Germany.

\end{document}